\documentclass[a4paper,10pt]{article}

\usepackage[english]{babel}
\usepackage[top=2.5cm, left=2.5cm, right=2cm, bottom=2.5cm]{geometry}
\usepackage{graphicx}
\usepackage[centertags]{amsmath}
\usepackage{amsfonts}
\usepackage{amssymb}
\usepackage{amsthm}
\usepackage{newlfont}

\newcommand{\beq}{\begin{equation}}
\newcommand{\eeq}{\end{equation}}
\newcommand{\bea}{\begin{eqnarray}}
\newcommand{\eea}{\end{eqnarray}}
\newcommand{\p}{\partial}
\newcommand{\nn}{\nonumber}
\begin{document}

\begin{center}

{\Large \bf{ The spin jumping in the context of a QCD effective model}}

\vspace{1cm}

G. B. de Gracia \\

\vspace{.5cm}

$^{1,2}$\emph{Instituto de F\'isica Te\'orica, UNESP - S\~ao Paulo State University},

\emph{P. O. Box 70532-2, 01156-970, S\~ao Paulo, SP, Brazil.\\
{\it{gb9950@gmail.com}} }

\end{center}

\vspace{.25cm}

\begin{abstract}

The tensor formulation for the effective theory of QCD vector ressonances, whose model we denote by TEVR, is given by an antisymmetric tensor field and describes spin $1$ particles. Our goal is to show, by diferent approaches, that the Abelian version of this model presents the so called ``spin jumping''  when we consider its massless limit. Classically we find, by the use of the equations of motion and the Hamiltonian constraint analysis, that the massive phase of the model describes  spin $1$  particles while its massless phase describes spin $0$ particles. By the quantum point of view we derive these conclusions via tree level unitarity analysis and the master action approach. 
PACS number:

\vspace{.5cm}

\end{abstract}

\section{Introduction}

\indent According to Ref. $1$ and $2$, we know that models described by antisymmetric fields present a discontinuity in its degrees of freedom, which we call spin jumping, when considering the massless limit. This is due to the fact that in the massless phase $p$ forms are dual to $D-p-2$ forms while in the massive phase the $p$ forms are dual to $D-p-1$ forms. There are also another cases where this discountinuity may occur, for example, near the massless limit of the topologically massive spin $1$ model of Ref. $3$ in which a four dimensional BF term, which do not have any local degrees of freedom Ref. $4$, becomes dominant. Alternatively, it may happen in some models that the massless equations of motion and the boundary conditions determines just a trivial solution to the fields while its massive phase possesses a definite spin content, as we can see in the appendix A. \\
\indent The main goal of this article is to verify the occurrence of this phenomena in the context of the Abelian version of the TEVR model, which is an effective theory for the vector ressonances of particular importance in QCD as we can see in Refs. $5$, $6$ and $7$. More specifically, it describes the ressonances $1^{--}$ which are excited states of quark bound states with spin $1$. So, for some circumstances it is easier to study these bound states as  pointwise particles via some effective action. We may mention the fact that due to the spin discountinuity of the model some approximations must be done carefully in order to have a consistent effective description. For example, in performing some calculations in the $S$ matrix for high momentum one often neglect the particle masses, but in the present context this approximation cannot be so radical since in this limit the TEVR model predict a different spin behaviour to the effective pointwise particle which is related, among other facts, to a change in its interparticle potential. In fact, this potential presents a DVZ discountinuity as we can see in Ref.$8$.\\
\indent The model is described by an antisymmetric field $B_{\mu \nu}(x)$ and its massive phase describes spin $1$ particles, while its massless phase describes spin $0$ excitations. An important motivation to this article is that a complete knowledge of the spin jumping phenomena may give us some insight to use this degree of freedom discountinuity in the context of an effective description for phase transitions controled by the  mass parameter of the TEVR model. It is also interesting to mention, as it was expected due the previous discussion, that the Kalb Ramond model, see Refs. $9$ and $10$, which also uses an antisymmetric field, presents the same kind of degree of freedom discontinuity studied in Ref. $1$. We explore this fact to do some general observations and analogies throughout this article.\\
\indent In order to analyze the spin jumping phenomena for the TEVR model we use diferent approaches:\\ Regarding the  Sec. $2$, the ``on shell" degrees of freedom of the model, in its massive and massless phases, are inferred by the  constraints of the equations of motion. In Sec. $3$, we present a counting of degrees of freedom for the massive/ massless phases of the model which is based on the Hamiltonian formalism, more specifically, the Dirac-Bergmann algorithm. The Sec. $4$ is devoted to the inference of  the `spin jumping" by the quantum point of view, through the tree level unitarity analysis. In Sec. $5$, the master action approach is used as a way to show the dual relation between the TEVR and the Maxwell-Proca  model while its massless phase can be related to a real scalar field model. This can be understood as an another method to infer the spin jumping. Finally, in Sec. $6$ we conclude. Technical details are relegated to the appendices.

\indent Natural units are used troughout  and the Minkowiski metric is $diag\ (-1,+1,..., +1)$.

\section{ Spin Jumping and the Equations of Motion }

\indent In this section we show that the equations of motion of the TEVR model can be used as a way to infer the ``on-shell" degrees of freedom propagated by this model in its massive and massless phases.\\

\subsection{The massive phase}

\indent The massive phase of the TEVR model is given by the action below:

\bea S=\int d^Dx\  \Bigg[ \p^\mu B_{\mu \nu}\ \p_\beta B^{\beta \nu}+\frac{m^2}{2}B_{\mu \nu}B^{\mu \nu} \Bigg] \eea

 Where  $B_{\mu \nu}=-B_{\nu \mu}$.\\
\indent The equations of motion can be obtained by the variational principle and can be rewritten in terms of the antisymmetric spin operators \footnote{Which are defined in the appendix.}:

\bea \Bigg[(m^2-\Box)P^{1e}_{\mu \nu \alpha \beta}+m^2\, P^{1b}_{\mu \nu \alpha \beta}\Bigg]B^{\mu \nu}(x)=0\eea

\indent By acting with $P^{1b}$ in the above equation and using the orthogonality of the antisymmetric spin operators we can show:

\bea m^2\,P^{1b}_{\mu \nu \alpha \beta}B^{\mu \nu}(x)=0\quad                             \eea

\indent The constraint above removes  $\frac{ (D-1)(D-2)}{2}$ degrees of freedom from the $B_{\mu \nu}$ field. This result allows us to conclude that an antisymmetric field that obeys this constraint propagates $D-1$ degrees of freedom, which is characteristic of a massive spin $1$ particle in $D$ dimensions.\\
\indent Now, we need to show that the $B_{\mu \nu}(x)$ field components obeys the Klein-Gordon equation. It can be done by using the identity\footnote{The symbol $I_{\mu \nu \alpha \beta}$ refers to the rank 4 antisymmetrized identity whose form lies in the apendix.} $P_{\mu \nu}^{\ \ (1e)\ \alpha \beta}=I_{\mu \nu}^{\ \  \alpha \beta}-P_{\mu \nu}^{\ \ (1b)\ \alpha \beta}$ and the equation $(3)$ applied in the equation $(2)$.\\

\subsection{The massless phase}
\indent The massless phase of the TEVR model presents a gauge symmetry in analogy to the massless Kalb Ramond theory. We could, at first, fix this symmetry to obtain the correct number of  its degrees of freedom. However, this is not the easiest way to do it. So, we follow a more direct approach which is given below.\\
\indent The equations of motion for the massless limit of the TEVR model can be written as:

\bea \Box P^{1e}_{\mu \nu \alpha \beta}B^{\mu \nu}(x)=0\eea

\indent The fact that this model, as its equations of motion, can be written in terms of just one spin operator leads us to conclude that it must have a gauge symmetry of the form \footnote{Square brackets denote antissymetrization.}:

\bea \delta B_{\mu \nu}(x)=\p^\alpha \Lambda_{\alpha \mu \nu}(x)\quad ; \quad \Lambda_{\alpha \mu \nu}(x)=\Lambda_{[\alpha \mu \nu]}(x)\eea

\indent In order to immediately infer the degrees of freedom propagated by the massless $B_{\mu \nu}$ field, we put aside the gauge symmetry discussion and focus on the informations contained in the equations of motion. 
\newpage
So, we use the fact that they can be rewritten in the following manner:

\bea \p^\mu(\p_\gamma B^{\gamma \nu})-\p^\nu(\p_\gamma B^{\gamma \mu})=0\eea

\indent From the above equations we can see that the divergence of the $B_{\mu \nu}$ field is the gradient of a scalar field $\p_\gamma B^{\gamma \nu}=\p^\nu \phi$. By contracting $B_{\mu \nu}$ with two derivatives and from its antisymmetry we can conclude:

\bea \p_\nu\p_\gamma B^{\gamma \nu}(x)=\p_\nu\p^\nu\phi(x)\to \Box \phi(x)=0\eea

\indent Therefore, the equations of motion of the massless TEVR model are equivalent to a harmonic scalar field in analogy to what happens to the massless Kalb Ramond model in $D=3+1$ dimensions according to Ref. $11$. Thus, we show, by the use of the equations of motion, that the spin jumping phenomena occurs, at least on the mass shell of the TEVR model. It is important to mention that the degree of freedom analogy between the TEVR and the Kalb Ramond model holds just in $D=3+1$ dimensions\footnote{This is due to the fact that the dynamical part of Kalb Ramond model is found in the space projected by the operator $P^{1b}_{\mu \nu \alpha \beta}$. So its massive phase has $\frac{ (D-1)(D-2)}{2}$ degrees of freedom, which is equivalent to spin $1$ just in $D=3+1$ dimensions. Its massless phase can be shown to be dual to a scalar field just in $D=3+1$ dimensions see Ref. $11$.}.\\

\section{Hamiltonian Analysis in D Dimensions}

\indent The Hamiltonian analysis, in particular the Dirac-Bergmann algorithm, see Refs. $12$ and $13$, provides us with a powerful method to obtain the degrees of freedom of some theory. To proceed with this analysis we divide it again in two parts, for the massive and for the massless phases of the theory. \\

\subsection{Hamiltonian analysis of the massive phase}

\indent The Lagragian density of the TEVR model is given below:

\bea {\cal{L}}=(\p^\mu B_{\mu \nu})^2+\frac{m^2}{2}B_{\mu \nu}B^{\mu \nu}\eea

Where we use the stardart notation $X_\mu X^\mu \equiv (X^{\mu})^2$.\\
\indent The canonical momenta follows from the expression\footnote{Where $\dot B_{\mu \nu}$ means $\p_0B_{\mu \nu}$.} $\pi^{\mu \nu}=\p{\cal{L}}/ \p \dot B_{\mu \nu}$. So, we have:

\bea \pi^{ij}=0\quad ; \quad \pi^{0i}=2(\dot B_{0i}+\p^jB_{ji})\eea

\indent From the definition of the canonical momenta we can obtain the primary constraints:

\bea \alpha^{ij}\equiv \pi^{ij}\thickapprox 0\eea

\indent The canonical Hamiltonian is given by the expression:

\bea {\cal{H}}_c=\pi^{0i}\dot B_{0i}-{\cal{L}}=\frac{(\pi^{0i})^2}{4}-\pi^{0i}\p^jB_{ji}+(\p^iB_{0i})^2-\frac{m^2}{2}(B_{ij})^2+m^2(B_{0j})^2\eea

\indent To obtain the primary Hamiltonian  we add Lagrange multipliers to each of the primary constraints:
  
\bea H_p=\int d^{D-1}x\,\Bigg[{\cal{H}}+\lambda_{ij}\alpha^{ij}\Bigg]\eea
\newpage
\indent To proceed with the Hamiltonian analysis it is necessary to evaluate the consistency of the constraints but to do so we first need to know the fundamental non-vanishing Poisson brackets of the theory:

\bea \{B_{\mu \nu}(x),\pi^{\alpha \beta}(y)\}=\delta^{(D-1)}{(x-y)}\frac{(\delta^\alpha_\mu \delta_\nu^\beta-\delta_\mu^\beta \delta_\nu^\alpha)}{2}\eea

\indent Now we may check the vanishing of the constraint's time evolution:

\bea \dot \alpha^{ij}(x)=\{\alpha^{ij}(x),H_c\}=\Bigg [\frac{F^{ij}(\pi^{0i})}{2}-m^2B^{ij}\Bigg]=0\eea

 Where we have that $F^{ij}(\pi^{0i})\equiv \p^i\pi^{0i}-\p^j\pi^{0j}$.

\indent In order to guarantee the consistency of the primary constraints we are lead to consider the secondary ones:

\bea \Phi^{ij}(x)\equiv \frac{F^{ij}(\pi^{0i})(x)}{2}-m^2B^{ij}(x)\thickapprox 0\eea

\indent The time evolution of the new constraints above are given by:

\bea \dot \Phi^{ij}(x)=\{\Phi^{ij}(x),H_c\}=2m^2F_{ij}(B^{0i})-m^2\lambda^{ij}= 0\eea

\indent We can conclude from the above result that the Dirac-Bergmann algorithm came to its end since the Lagrange multipliers are determined. It is not dificult to check that all the constraints are of second class which is related to the fact that the massive TEVR do not have gauge symmetry (The Dirac  brackets of the system are given in the appendix C).\\
\indent Regarding the counting of degrees of freedom, we had initially $D(D-1)$ phase space degrees of freedom, namely the fields and its canonical momenta. When we take the constraints into account we have that the primary ones $\alpha^{ij}(x)$ remove $\frac{(D-1)(D-2)}{2}$  degrees of freedom, while the secondary ones remove this same quantity. Thus, in the end of the day we have $2(D-1)$ phase space degrees of freedom which are compatible to a spin $1$ particle in $D$ dimensions in according to the previous section\footnote{We must remember that the phase space has double the dimensions of the configuration space.}.\\
\indent Considering the constraints as strong equalities the Hamiltonian become positive definite which guarantees the classical stability of the model:

\bea H_c=\int d^3x\ \Bigg[\frac{(\pi^{0i})^2}{4}+(\p^iB_{0i})^2+\frac{m^2}{2}(B_{ij})^2+m^2(B_{0j})^2\Bigg]\geq 0\eea\\

\subsection{Hamiltonian analysis of the massless phase}

\indent The Lagragian density of the massless phase of the TEVR model is given by:

\bea {\cal{L}}=(\p^\mu B_{\mu \nu})^2\eea

\indent Since the kinetic part of the massless phase of the model is the same that of the massive phase their primary constraints must be the same. The Hamiltonian density is given by the following expression:

\bea {\cal{H}}_c=\frac{(\pi^{0i})^2}{4}-\pi^{0i}\p^jB_{ji}+(\p^iB_{0i})^2\eea

\indent The primary Hamiltonian is obtained by associating Lagrange multipliers to each of the constraints:

\bea H_p=\int d^{D-1}x\ \Bigg[{\cal{H}}_c+\lambda_{ij}\alpha^{ij}\Bigg]\eea

\indent The first difference between the constraint structure of the massive and the massless versions of the TEVR models lies on its primary constraints consistency conditions:

\bea  \dot \alpha^{ij}(x)=\{\alpha^{ij}(x),H_c\}=F^{ij}(\pi^{0i})=0\eea

\indent From the above expression we conclude that secondary constraints are needed:

\bea \Omega^{ij}\equiv F^{ij}(\pi^{0i})\thickapprox 0\eea

\indent The analysis of those new constraints is more intricate than the previous ones. The reason is due to the fact that we are lead to the issue of reducibility, see  Ref. $21$, due to the existence of the relations \footnote{$ P^{1b}_{ij mn}$ is the spatial part of the antisymmetric spin projector. See appendix B}:

\bea P^{1b}_{ij mn}\Omega^{mn}=0\eea

The presence of reducibility is related to the fact that to quantize the system, one must introduce extra ghost terms in the gauge fixed Lagrangian in order to have a system without any local freedom.\\
\indent We could, at principle, perform a Hamiltonian analysis inside the context of reducibility but we find it simpler to follow the alternative way of deriving a general solution to the constraints $(22)$. So, in the constraint surface we have:

\bea \pi^{0i}(x)=\p^i\gamma(x)\eea

Where $\gamma(x)$ is a scalar field with an appropriate mass dimension.\\
\indent The  $\Omega^{ij}$ constraint forces $\pi^{0i}$ to be purely longitudinal. So it removes $D-2$ degrees of freedom from the model, a diferent quantity that is removed by the secondary constraints in the massive phase.\\
\indent The consistency conditions are identically satisfied:

\bea \dot \Omega^{ij}(x)=\{\Omega^{ij}(x),H_c\}=0\eea

\indent Thus, the algorithm comes to its end and we can infer that there is the presence of gauge symmetry since the constraints are now of the first class and the Lagrange multipliers are indetermined which means that the theory has some arbitrary local freedom.\\
\indent The degree of freedom count is done by removing from the initial $D(D-1)$ dimensional phase space two degrees of freedom for each first class constraint. The necessity for removing the double of degrees of freedom that would be naively removed by the constraints is due to the fact that for each first class constraint we should fix one corresponding gauge fixing condition to the field $B_{\mu \nu}(x)$ in order to have a uniquely determined dynamics for it.\\
\indent Regarding the counting of the degrees of freedom we have in the end of the day two  phase space degrees of freedom, which corresponds to just one in the configuration space, so the massless limit of the TEVR model describes a spin $0$ particle.\\
\indent The classical stability of the theory is guaranteed by the fact that in the constraint surface the Hamiltonian is positive definite:

\bea {\cal{H}}_c=\frac{(\pi^{0i})^2}{4}+(\p^iB_{0i})^2 \geq 0\eea

\indent From the Hamiltonian analyisis we can recover the results obtained by the equations of motion through a more rigorous way. Our next step is to show that at the quantum level we can also infer this degree of freedom discontinuity when considering the massless limit.\\

\section{Tree Level Unitarity Analysis }

\indent In this section we show that by means of the unitarity analysis, see Refs. $14$ and $15$, we can infer the physical degrees of freedom of the TEVR model by identifying the form of the terms that contributes to the residue of the saturated amplitude. First we will treat the massive and then its massless phase to show the occurrence of the spin jumping under the quantum point of view. The procedure of verifying the tree level unitarity consists in coupling an external classical source to the fields of the free Lagrangian. This interaction term will generate contributions to the functional generator which can be expressed by means of tree diagrams with this source attached to the external lines. Since this source is weakly coupled to the theory, in order to verify its unitarity it is enough to look at the positivity of the  imaginary part of its first non trivial contribution. In practice it can be done by using  a simple application of the Cutkosky rules.

\subsection{On the unitarity of the massive phase}

\indent In order to perform the unitarity analysis of TEVR model it is useful to express its action  in terms of the antisymmetric spin operators and also add a source term:

\begin{multline} S=\int d^Dx\,\Bigg[(\p^\mu B_{\mu \nu})^2+\frac{m^2}{2}B_{\mu \nu}B^{\mu \nu}+B_{\mu \nu}T^{\mu \nu}\Bigg ]=\int d^Dx\, \bigg \{\frac{\ B^{\mu \nu}}{2}\Bigg[(m^2-\Box)P^{1e}_{\mu \nu \alpha \beta}+m^2P^{1b}_{\mu \nu \alpha \beta}\Bigg]\,B^{\alpha \beta}+B_{\mu \nu}T^{\mu \nu}\bigg \}\end{multline}
\indent Where $T_{\mu \nu}=-T_{ \nu \mu}$.\\
\indent The action differential operator, expressed in terms of the spin operators, reads:

\bea \hat O_{\mu \nu \alpha \beta}=\frac{[(m^2-\Box)P^{1e}_{\mu \nu \alpha \beta}+m^2P^{1b}_{\mu \nu \alpha \beta}]}{2}\eea

\indent The saturated amplitude can be obtained by contracting the inverse of this differential operator with the antisymmetric sources. In the momentum space this calculation becomes straightforward. So, the amplitude is given by:

\bea {\cal{A}}_{(m\neq 0)}(k)=-\frac{i}{2}{T^{\mu \nu}}^*(k)\hat O_{\mu \nu \alpha \beta}^{-1}(k)T^{\alpha \beta}(k)=-i{T^{\mu \nu}}^*(k)[\frac{P^{1e}_{\mu \nu \alpha \beta}}{(m^2+k^2)}+                                                                                                                     \frac{P^{1b}_{\mu \nu \alpha \beta}}{m^2}]T^{\alpha \beta}(k) \eea

\bea {\cal{A}}_{(m\neq 0)}(k)=-i[\frac{-\vert k^\mu T_{\mu \nu}\vert^2}{m^2(k^2+m^2)}+\frac{\vert T_{\mu \nu}\vert^2}{m^2}]\eea

\indent The unitarity analysis is done by verifying the signal of the imaginary part of the residue of the saturated amplitude. From the above  expression we can conclude that the term that contributes to the residue comes from the contribution of the operator $P^{1e}_{\mu \nu \alpha \beta}$. This operator has the property of projecting in a $D-1$ dimensional space:

\bea \Im \lim_{k^2\to -m^2}(k^2+m^2)\,{\cal{A}}_{(m\neq 0)}(k)=\frac{\vert k^\mu T_{\mu \nu}\vert^2}{m^2}\eea

\indent In order to show that the above quantity is positive definite, which means that the theory is unitary, we use the definition:

\bea \vert k^\mu T_{\mu \nu}(k) \vert \equiv J_\nu^T(k)\eea

\indent Where the index $`` T "$ designates transversality.\\
\indent A convenient frame to perform the calculation of $(31)$ is given below:

\bea k_\mu=(m,0,0,..., 0)\quad ; \quad k^\mu J_\mu^T(k)=0\quad ; \quad J_0^T=0\eea

\indent In this frame we can use the transversality of $J_\mu^T(k)$ to show that $J_0^T=0$. Thus, the expression of $(31)$ becomes positive definite and we can infer that the model is tree level unitary:

\bea \Im \lim_{k^2\to -m^2}(k^2+m^2)\,{\cal{A}}_{(m\neq 0)}(k)>0\eea

\indent From this result we can conclude that the contribution to the unitarity comes from the $P^{1e}$ spin operator which projects the tensor field in its spin $1$ sector. This result is in accordance to our earlier results which were obtained by classical methods.\\

\subsection{On the unitarity of the Massless phase}
\indent The unitarity analysis of the massless phase of the TEVR model is more intricate since its differential operator is not inversible due to its gauge freedom:

\bea \delta B_{\mu \nu}=\p^\alpha \Lambda_{[\alpha \mu \nu]}\eea

\indent Thus, in order to proceed with such an analysis it is necessary to add a gauge fixing term which will allow us to invert the diferential operartor. The gauge condition we choose is:

\bea G_{\lambda \mu \nu}=\p_\lambda B_{\mu \nu}+\p_\mu B_{\nu \lambda}+\p_\nu B_{\lambda \mu}=0\eea

\indent The fact that the system has gauge symmetry is related to constraints in the sources. Those constraints can be obtained by the requiring that the source term of the model must be invariant by the same symmetry transformations that leaves its quadratic part invariant:

\bea \int \delta {\cal{L}}_{source}d^Dx=\int \delta B_{\mu \nu}T^{\mu \nu}d^Dx=0 \to T_{\mu \nu}(k)=k_\mu J_\nu-k_\nu J_{\mu}\eea

\indent The gauge fixing term which is totally projected on the $P^{1b}$ operator has the form:
\bea {\cal{L}}_{g.f}=\frac{\lambda}{2} \,G_{\lambda \mu \nu}(B)G^{\lambda \mu \nu}(B)\eea

\indent The gauge fixed Lagrangian reads:

\bea {\cal{L}}=(\p^\mu B_{\mu \nu})^2+\frac{\lambda}{2} \,G_{\lambda \mu \nu}(B)G^{\lambda \mu \nu}(B)\eea

\indent By using the form $(37)$ for the sources and the same procedure adopted in the previous subsection we obtain the saturated amplitude:

\bea {\cal{A}}_{(m=0)}(k)=-i\frac{{T^{\mu \nu}}^*(k)P^{1e}_{\mu \nu \alpha \beta}T^{\alpha \beta}(k)}{k^2}=-i\frac{(k^4\vert j_\beta \vert^2-k^2\vert k^\alpha J_\alpha\vert^2)}{k^4} \eea

\indent From the above expression we have that: 

\bea \Im \lim_{k^2\to 0}k^2{\cal{A}}_{(m=0)}(k)>0\eea

\indent This result allow us to conclude that the massless phase of the TEVR model is unitary at tree level . Regarding the degrees of freedom we note that the contribution to the residue of the saturated amplitude comes from a term of the form $\vert k^\alpha J_\alpha\vert^2$ which can be written as $k^2{j^\nu}^* \omega_{\nu \mu}J^\mu$ with $\omega_{\nu \mu}$ being the vector spin $0$ operator  (see apendix B.).\\
\indent Therefore, we note that the massless phase contribution to unitarity comes from a term that can be expressed  by the vector spin $0$ operator. This result is in accordance with the analysis of the previous sections and when compared to the Ref. $10$ it can be understanded as a quantum level spin jumping.\\

\section{Master Action and Degrees of Freedom}

\indent The goal of this section is to obtain a master action, see Refs. $16$ and $17$, that relates the TEVR with the Maxwell-Proca above  in $D$ dimensions and one that relates its massless limit with the action of a scalar field.\\
\indent The relation between the TEVR  and the Maxwell-Proca model is already inferred, but under diferent approaches as can be seen in Ref. $5$. Our goal is to show that this relation can be also obtained by using a master action. From these results we can finally compare them to the ones obtained for its massless phase and find, now by the master action approach, the already mentioned degree of freedom discontinuity.\\
\indent By adding sources to the master action fields it is possible, for example, to compare its two point functions and obtain a dual map between then. In the massless phase we show that there is a problem in the determination of this dual map, a fact that has an direct analogue when considering the massless Kalb Ramond model.\\

\subsection{Massive phase and the Maxwell-Proca model}

\indent The master action that interpolates the TEVR and the Maxwell-Proca model has the form: 

\bea S_M^{(m\neq 0)}=\int d^Dx\ \Bigg[ m^2B_{\mu \nu}B^{\mu \nu}+2m(\p^\mu B_{\mu \nu})A^\nu-\frac{m^2}{2}A^\mu A_\mu+A_\nu J^\nu+B_{\mu \nu}T^{\mu \nu}\Bigg]\eea

\indent A Gaussian integration in the vector field $A_\mu(x)$ lead us to:

\bea S_M^{(m\neq 0)}=\int d^Dx\Bigg[2\,(\p^\mu B_{\mu \nu})^2+\frac{J_\nu \, J^\nu}{2m^2}+\frac{2\,(\p^\mu B_{\mu \nu})J^\nu}{m}+m^2B_{\mu \nu}B^{\mu \nu}+B_{\mu \nu}T^{\mu \nu}\Bigg]\eea

\indent The action above is the massive TEVR model coupled to source terms. To verify this result we redefine the fields  $B_{\mu \nu}(x) \to \frac{\tilde B_{\mu \nu}(x)}{\sqrt{2}}$.\\
\indent In order to show that the Maxwell-Proca theory can be related to the TEVR we use the expression $(42)$ to perform a Gaussian integration in $B_{\mu \nu}(x)$:

\bea S_M^{(m\neq 0)}=\int d^Dx\, \Bigg[-\frac{1}{4}F_{\mu \nu}F^{\mu \nu}-\frac{T_{\mu \nu}T^{\mu \nu}}{4m^2}+\frac{F^{\mu \nu}T_{\mu \nu}}{2m}-\frac{m^2}{2}A_\nu A^\nu+A_\nu J^\nu \Bigg ]\eea

\indent Where $F^{\mu \nu}(A)\equiv\p^\mu A^\nu-\p^\nu A^\mu$.\\
\indent The fact that the Maxwell-Proca and the TEVR model can be obtained from the same master action lead us to infer that they must have a common particle content. Both models describes spin $1$ particles.\\
\indent The dual map between those models can be obtained by considering their two point functions. In order to calculate them is necessary to use the functional generators:

\bea Z_M^{(1)(m\neq 0)}[T_{\mu \nu},J_\nu]=\int {\cal {D}}\tilde B_{\mu \nu}(x)e^{i\int d^Dx\,\Bigg[\,(\p^\mu \tilde B_{\mu \nu})^2+\frac{J_\nu \, J^\nu}{2m^2}+\frac{\sqrt{2}\,(\p^\mu \tilde B_{\mu \nu})J^\nu}{m}+\frac{m^2}{2}\tilde B_{\mu \nu}\tilde B^{\mu \nu}+\frac{\tilde B_{\mu \nu}}{\sqrt{2}}T^{\mu \nu}\Bigg]}\eea

\bea Z_M^{(2)(m\neq 0)}[T_{\mu \nu},J_\nu]=\int {\cal {D}}A_{\mu}(x)e^{i\int d^Dx\Bigg[-\frac{1}{4}F_{\mu \nu}F^{\mu \nu}-\frac{T_{\mu \nu}T^{\mu \nu}}{4m^2}+\frac{F^{\mu \nu}T_{\mu \nu}}{2m}-\frac{m^2}{2}A_\nu A^\nu+A_\nu J^\nu \Bigg]}\eea
\indent The first of the above functional generators is obtained from $ S_M^{(m\neq 0)}$ through a Gaussian integration in the field $A_\mu(x)$ while the second one comes from an integration in the $B_{\mu \nu}(x)$ field. By performing functional derivatives we get the two point functions:

\bea \frac{1}{2}<\tilde B_{\mu \nu}(x)\tilde B_{\alpha \beta}(y)>=-\frac{\delta^2Z_M^{(1)(m\neq 0)}[T_{\mu \nu},J_\nu]}{Z_M^{(1)(m\neq 0)}[0,0]\ \delta \, T_{\mu \nu}(x) \, \delta \, T_{\alpha \beta}(y)}\Bigg \vert_{T_{\alpha \beta}=J_\nu=0}\eea

\indent If we vary $Z_M^{(2)(m\neq 0)}[T_{\mu \nu},J_\nu]$ with relation to the tensor sources, we should obtain a result that describes the same physics than the one obtained above since 
$Z_M^{(1)(m\neq 0)}[T_{\mu \nu},J_\nu]$ as well $Z_M^{(2)(m\neq 0)}[T_{\mu \nu},J_\nu]$ are originated by functional integrations of the same master action.\\
\indent So, the two point function is given by the expression below:

\bea -\frac{\delta^2Z_M^{(2)(m\neq 0)}[T_{\mu \nu},J_\nu]}{Z_M^{(2)(m\neq 0)}[0,0]\,\delta \, T_{\mu \nu}(x) \, \delta \, T_{\alpha \beta}(y)}\Bigg \vert_{T_{\alpha \beta}=J_\nu=0}=\frac{1}{4m^2}<F^{\mu \nu}(x)F^{\alpha \beta}(y)>+\frac{i}{2\,m^2}\delta^{(D-1)}(x-y)\ \quad \quad \eea

\indent From this last result we can obtain a dual map that relates the fields $B_{\mu \nu}(x)$ and $A_\mu(x)$ up to contact terms:

\bea  \tilde B_{\mu \nu}\longleftrightarrow \frac{\sqrt{2}\,F_{\mu \nu}(A)}{2\, m}\quad \quad \eea

\indent The inverse dual map can be obtained using the same argument employed in the above calculations. The difference lies in the fact that now we perform functional derivatives with respect to the vector sources.
The two point function is given by:

\bea <A_\mu(x)A_\nu(y)>=-\frac{\delta^2Z_M^{(2)(m\neq 0)}[T_{\mu \nu},J_\nu]}{Z_M^{(1)(m\neq 0)}[0,0]\, \delta \, J^{\mu}(x) \, \delta\, J^\nu (y)}\Bigg \vert_{T_{\alpha \beta}=J_\nu=0}\eea

\indent When we vary $Z_M^{(1)(m\neq 0)}$ with relation to the vector sources we should find a result that describes the same physics than the two point function calculated previously:

\bea -\frac{\delta^2Z_M^{(1)(m\neq 0)}[T_{\mu \nu},J_\nu]}{Z_M^{(1)(m\neq 0)}[0,0]\, \delta \, J^{\mu}(x) \, \delta \, J^\nu (y)}\Bigg \vert_{T_{\alpha \beta}=J_\nu=0}= -i\frac{\,\delta^{(D-1)}(x-y)}{m^2}+\frac{2}{m^2}<\p^\alpha \tilde B_{\alpha \mu}(x)\p^\gamma \tilde B_{\gamma \nu}(y)>\nonumber \\ \eea

\indent This result allow us to obtain the inverse dual map that relates the fields $A_\mu(x)$ and $B_{\alpha \mu}(x)$ up to contact terms:

\bea A_\mu(x)\longleftrightarrow \frac{\sqrt{2}\, \p^\alpha \tilde B_{\alpha \mu}}{m}\eea

 \indent The results obtained in this section are in agreement to the ones obtained in the Ref. $1$ and represent an alternative way to relate the Maxwell-Proca model, which describes a spin $1$ particle, to the massive TEVR model. This fact lead us to infer that it must describe a particle with this spin.\\

\subsection{Massless phase and a real scalar field model}

\indent We can show that the massless phase of the TEVR model can be related to a spin $0$ field by the use of the following master action:

\bea S_M^{(m=0)}=\int d^Dx\,\Bigg[-C_\nu C^\nu+2C^\nu(\p^\mu B_{\mu \nu})+C_\nu J^\nu+B_{\mu \nu}T^{\mu \nu}\Bigg ]\eea

\indent A Gaussian integration in the vector field lead us to the massless limit of the TEVR model coupled to sources:

\bea S_M^{(m=0)}=\int d^Dx\,\Bigg[(\p^\mu B_{\mu \nu})^2+\frac{J_\nu J^\nu}{4}+\p^\mu B_{\mu \nu}J^\nu+B_{\mu \nu}T^{\mu \nu}\Bigg ]\eea

\indent On the other hand, the master action $(53)$ can be rewritten in an useful way by means of an integration by parts: 

\bea S_M^{(m=0)}=\int d^Dx\,\Bigg [-C_\nu C^\nu-F^{\mu \nu}_cB_{\mu \nu}+C_\nu J^\nu+B_{\mu \nu}T^{\mu \nu}\Bigg ]\eea

\indent Where $F^{\mu \nu}_c(x)\equiv \p^\mu C^\nu(x)-\p^\nu C^{\mu}(x)$.\\

\indent The integration in the $B_{\mu \nu}(x)$ field is equivalent to integrating a functional delta function which imposes $F^{\mu \nu}_c=T^{\mu \nu}$. The solution is given below:

\bea C_\nu=\p_\nu \psi+C_\nu^T\eea

\indent The vector field decomposes in a longitunal part, which is the solution of the homogeneous equation and in a tranverse part which is given by:

 \bea C_\nu^T(x)=\frac{\p^\mu T_{\mu \nu}}{\p^2}\eea

\indent If we plug the above result in the action $(55)$ we find:

\bea S_M=\int d^Dx\,\Bigg[-\p^\nu \psi \, \p_\nu \psi+\p_\nu \psi \,J^\nu-{\Bigg (\frac{\p^\mu T_{\mu \nu}}{\p^2}\Bigg)}^2+\frac{\p^\mu T_{\mu \nu}\,J^\nu}{\p^2}\Bigg ] \eea

\indent The result obtained above shows that the massless phase of the TEVR model, up to source terms, is equivalent to the action of a real scalar field which describes a spin $0$ particle. So, considering the results from the previous subsection we conclude that the master action technique can be used as another way  for inferring the ocurrence of the spin jumping.\\
\indent The functional generators are given by:

\bea Z_M^{(1)(m=0)}[T_{\mu \nu},J_\nu]=\int dB_{\mu \nu}(x)e^{i\int d^Dx\Bigg[(\p^\mu B_{\mu \nu})^2+\frac{J_\nu J^\nu}{4}+\p^\mu B_{\mu \nu}J^\nu+B_{\mu \nu}T^{\mu \nu}+\frac{\lambda}{2} \,G_{\lambda \mu \nu}(B)G^{\lambda \mu \nu}(B)\Bigg]}\eea

\bea Z_M^{(2)(m=0)}[T_{\mu \nu},J_\nu]=\int d\,\psi(x)\, e^{i\int d^Dx\Bigg[-\p^\nu \psi \, \p_\nu \psi+\p_\nu \psi \,J^\nu-{(\frac{\p^\mu T_{\mu \nu}}{\p^2})}^2+\frac{\p^\mu T_{\mu \nu}}{\p^2}J^\nu \Bigg]}\eea

\indent In the above expression we add a gauge fixing term because this action has the symmetry mentioned in $(35)$.

\indent A peculiarity that is present in the massless limit of the TEVR model is related to the obtainment of the dual map. Since there is no interaction terms that are linear in the tensor sources in $Z_M^{(2)(m=0)}$ we cannot find an inversible map by the same procedure adopted in the previous subsection. So, we restrict ourselves to infer that a scalar spin $0$ field action and the massless TEVR are equivalent up to source terms. This result is again in accordance to our earlier results obtained by diferent approaches.\\
\indent Although we cannot obtain an invertible dual map one can find, by the same procedure adopted in the previous section, a mapping between vector two point functions:

\bea     <\p^\nu \psi(x)\p^\mu \psi(y)> =<\p_\beta B^{\beta \nu}(x) \p_\omega B^{\omega \mu}(y)>-i\frac{\delta^{(D-1)}(x-y)}{2} \eea

\indent The above result is clearly in accordance with the one given in the end of the Sec. $2$ up to contact terms. Although this statement relates classical and quantum results it is not surprising since our calculations are being performed at tree level.

\section{Concluding Remarks}

\indent This article consisted basically in verifying the spin jumping phenomena for the TEVR model  under different approaches. Regarding the classical analyisis, in the Sec. $2$  we found this phenomena by means of the equations of motion while in the Sec. $3$ we used the Hamiltonian formalism and found that its massless phase has reducible constraints as well as all the discountinuous models cited in this paper. \\
\indent We could also understand this degree of freedom discountinuity under a quantum approach via unitarity analysis and master action technique presented in the Secs. $4$ and $5$, respectivelly. \\
\indent As a future perspective we intend to present a phenomenological application of this degree of freedom discountinuity in the context of phase transitions with the mass as a parameter. To do so, we should derive its partition function and understand how this spin discountinuity appears in this context. So, regarding the massless phase, it is also necessary to have a clear picture of the role played by the constraint's reducibility in the construction of its gauge fixed Lagrangian.

\section{Acknowledgements}

\indent We thank the referee for the interesting suggestions. This work was supported by CNPq (132619/2015-6).

\section{Appendix A: A discountinuous vector model}

Consider the spin $0$ model described by the longitudinal excitations of a vector field:
\bea {\cal{L}}=\frac{1}{2}\bigg[(\p_\mu C^\mu)^2+m^2C_\mu C^\mu \bigg]\eea

\indent The equations of motion (E.O.M) are given by:

\bea -\p_\mu (\p_\nu C^\nu)+m^2C_\mu=0\eea

\indent The spin content of the theory can be obtained by contracting the above equation with a derivative:

\bea (\Box-m^2)(\p_\mu C^\mu)=0\eea

\indent It is clear that we have massive scalar excitations described by $\p_\mu C^\mu$. On the other hand the massless model presents a local freedom, which is analogous to the massless TEVR one:

\bea \delta C_\mu(x)=\p^\nu \Gamma_{[\nu \mu]}(x)\eea
Where $\Gamma_{[\nu \mu]}(x)$ is an arbitrary antissimetric field. The local transformation above is reducible since it is invariant by $\delta \Gamma_{[\nu \mu]}(x) \to \bar \Gamma_{[\nu \mu]}^T(x)$ where $T$ designates transversality. The E.O.M are:
\bea \p_\mu (\p_\nu C^\nu)=0\eea
This equation determines that $\p_\nu C^\nu$ is a constant. So, requiring that the fields falls to zero at spacetime infinity force us to consider the trivial solution. The model becomes trivial in the massless limit. It can be understanded by the vanishing of its interpaticle potential in this limit (Ref.$8$).

\section{Appendix B: Spin projectors}

 Using the spin-0 and spin-1 projection
ope

rators acting on vector fields, respectively,

\bea  \omega_{\mu\nu} = \frac{\p_{\mu}\p_{\nu}}{\Box} \ , \
\theta_{\mu\nu} = \eta_{\mu\nu} -
\frac{\p_{\mu}\p_{\nu}}{\Box}\quad , \label{pvectors} \eea

 as building blocks, one can define the projection
 operators in $D$ dimensions acting on antisymmetric rank-2
tensors.
First, we define the transversal and longitudinal operators as follows
\begin{eqnarray}
\theta_{\mu\nu} = \eta_{\mu\nu} - \frac{\p_\mu \p_\nu}{\Box} \quad  \quad \omega_{\mu\nu} = \frac{\p_\mu \p_\nu}{\Box}.
\end{eqnarray}
The above set of operator satisfies 
\begin{eqnarray}
\theta^2 = \theta , \quad \omega^2 = \omega \quad \textmd{and} \quad \theta \omega = \omega \theta = 0. 
\end{eqnarray}

On the other hand, the set of the antisymmetric four-dimensional Barnes-Rivers operator are given by  
\begin{equation}
P^{[1b]}_{\mu\nu,\alpha\lambda} = \frac{1}{2}(\theta_{\mu\alpha} \theta_{\nu\lambda} - \theta_{\mu\lambda} \theta_{\nu\alpha}),
\end{equation}
\begin{equation}
P^{[1e]}_{\mu\nu,\alpha\lambda} = \frac{1}{2}(\theta_{\mu\alpha} \omega_{\nu\lambda} + \theta_{\nu\lambda} \omega_{\mu\alpha} - \theta_{\mu\lambda} \omega_{\nu\alpha} - \theta_{\nu\alpha} \omega_{\mu\lambda}).
\end{equation}
They satisfy the very simple algebra
\begin{eqnarray}
&(P^{[1b]})^2 = P^{[1b]}, \quad (P^{[1e]})^2 = P^{[1e]},&\nn\\
&P^{[1b]}P^{[1e]} = P^{[1e]}P^{[1b]}=0.
\end{eqnarray}

\section{Appendix C: The Dirac brackets of the massive TEVR model }

\indent The massive phase of the TEVR model has just second class constraints and it is related to the fact that the Lagrange multipliers are determined. So, according to the Dirac-Bergmann procedure we need to consider a reduced phase space given by the Dirac brackets. First of all, its necessary to build the constraint matrix:

\bea \Gamma^{ij}_{\ mn}(x,y)=\left(\begin{array}{ccccc}
\{\Omega^{ij}(x),\Omega_{mn}(y)\}&  \{      \Omega^{ij}(x),\alpha_{mn}(y)\} \\      
 \{   \alpha^{ij}(x),   \Omega_{mn}(y)\}& \{ \alpha^{ij}(x),\alpha_{mn}(y)\}    \\
  
\end{array}\right)= m^2\left(\begin{array}{ccccc}
0 & -1   \\
 1 & 0    \\
   
\end{array}\right) \frac{(\delta^i_m \delta^j_n-\delta^j_m \delta^i_n)}{2} \delta^{(D-1)}(x-y)    \quad               \eea

\indent The inverse of $ \Gamma^{ij}_{\ mn}(x,y)$  is given by:

\bea {\Gamma^{ij}_{\ mn}(x,y)}^{-1}=\frac{1}{m^2}\left(\begin{array}{ccccc}
0 & 1   \\
 -1 & 0    \\
   
\end{array}\right) \frac{(\delta^i_m \delta^j_n-\delta^j_m \delta^i_n)}{2} \delta^{(D-1)}(x-y)                                                                     \eea

\indent Now we can obtain the Dirac brackets that generates the reduced phase space of the system:

\bea \{A(x),B(y)\}_D=\{A(x),B(y)\}-\int d^{D-1}u d^{D-1}v\{A(x), \chi^a_{\ ij}(u)\}{\Gamma^{ij}_{\ mn\ (ab)}(u,v)}^{-1}\{\chi^{b\ mn}(v),B(y)\}\eea

Where $\chi^a_{\ ij}(u)=\left(\begin{array}{ccccc}
\Omega_{ij}(u)    \\
 \alpha_{ij}(u)    \\
   
\end{array}\right)$. The reduced phase space is given by the nonvanishing Dirac brackets:

\bea \{ B^{0i}(x),\pi_{0j}(y)\}_D=\frac{\delta^i_j}{2}\delta^{(D-1)}(x-y)\eea

\bea \{ B^{ij}(x),B_{0l}(y)\}_D=\frac{\p^{[j}\delta^{i]}_{\ l}}{2m^2}          \delta^{(D-1)}(x-y)\eea

Where $[\ , \ ]$ denotes antisymmetrization.

\end{document}